\documentclass[11pt]{article}
\usepackage{fullpage,latexsym,epsfig,amssymb,amsmath}

\newcommand{\adj}[1]{{#1}^{\dagger}}
\newcommand{\tuple}[1]{\mathord{\left\langle {#1} \right\rangle}}

\newcommand{\ket}[1]{\mathord{\left| {#1} \right\rangle}}

\newcommand{\braopket}[3]{\tuple{{#1}|{#2}|{#3}}}

\renewcommand{\phi}{\varphi}
\newcommand{\map}[3]{{#1} : {#2} \rightarrow {#3}}
\newcommand{\cH}{{\cal H}}

\newcommand{\eqdf}{\mathrel{:=}}

\newcommand{\setof}[1]{\left\{{#1}\right\}}
\newcommand{\two}{{\setof{0,1}}}
\newcommand{\condition}{\mathrel{:}}

\newcommand{\card}[1]{{\mathopen{\parallel}{#1}\mathclose{\parallel}}}
\newcommand{\CNOT}{{\rm CNOT}}

\newcommand{\Xtot}[1]{{\Sigma X}}
\newcommand{\Ytot}[1]{{\Sigma Y}}
\newcommand{\Ztot}[1]{{\Sigma Z}}
\newcommand{\Mod}{{\rm Mod}}
\newcommand{\diag}[2]{{{\rm diag}\left[\begin{array}{#1}#2\end{array}\right]}}

\title{Implementing the fanout gate by a Hamiltonian}

\author{Stephen A. Fenner\thanks{Computer Science and Engineering
Department, Columbia, SC 29208 USA.  Email {\tt fenner@cse.sc.edu}.
This work was supported in part by the National Security Agency
(NSA) and Advanced Research and Development Activity (ARDA) under Army
Research Office (ARO) contract number DAAD 190210048.} \\
University of South Carolina}

\date{\today}

\begin{document}

\bibliographystyle{hplain}

\maketitle

\begin{abstract}
We show that, for even $n$, evolving $n$ qubits according to a simple
Hamiltonian can be used to exactly implement an $(n+1)$-qubit parity
gate, which is equivalent in constant depth to an $(n+1)$-qubit fanout
gate.  We also observe that evolving the Hamiltonian for three qubits
results in an inversion-on-three-way-equality gate, which together
with single-qubit operations is universal for quantum computation.
\end{abstract}

\section{Introduction}

\begin{figure}
\begin{center}
\begin{picture}(0,0)%
\includegraphics{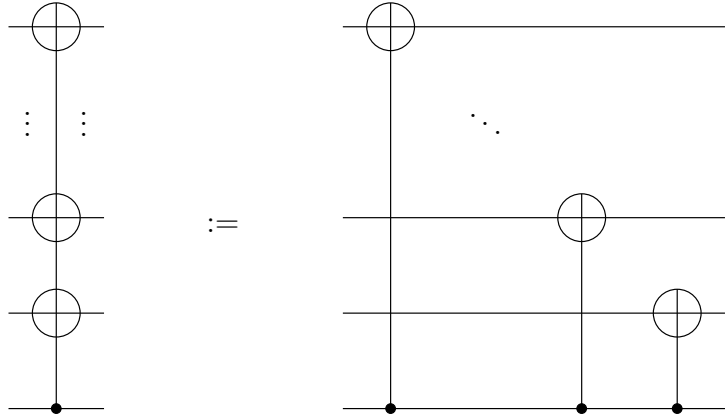}%
\end{picture}%
\setlength{\unitlength}{3947sp}%
\begingroup\makeatletter\ifx\SetFigFont\undefined%
\gdef\SetFigFont#1#2#3#4#5{%
  \reset@font\fontsize{#1}{#2pt}%
  \fontfamily{#3}\fontseries{#4}\fontshape{#5}%
  \selectfont}%
\fi\endgroup%
\begin{picture}(4524,2599)(2989,-3998)
\put(4351,-2836){\makebox(0,0)[b]{\smash{\SetFigFont{12}{14.4}{\rmdefault}{\mddefault}{\updefault}$:=$}}}
\put(3151,-2236){\makebox(0,0)[rb]{\smash{\SetFigFont{12}{14.4}{\rmdefault}{\mddefault}{\updefault}$\vdots$}}}
\put(3451,-2236){\makebox(0,0)[lb]{\smash{\SetFigFont{12}{14.4}{\rmdefault}{\mddefault}{\updefault}$\vdots$}}}
\put(6001,-2236){\makebox(0,0)[b]{\smash{\SetFigFont{12}{14.4}{\rmdefault}{\mddefault}{\updefault}$\ddots$}}}
\end{picture}
\caption{Definition of the fanout gate.}\label{fig:fanout-def}
\end{center}
\end{figure}

Let $\cH$ be the Hilbert space of $n+1$ qubits.  The \emph{fanout
operator} $\map{F_{n+1}}{\cH}{\cH}$, depicted in
Figure~\ref{fig:fanout-def},
copies the (classical) value of a single qubit to $n$ other qubits.

Fanout gates have been recently shown to be very powerful primitives
for making shallow quantum circuits
\cite{GHMP:fanout,HS:fanout,Moore:fanout,Spalek:fanout}.  It has been
shown that in the quantum realm, fanout, parity (see below), and
$\Mod_q$ gates (for any $q \geq 2$) are all equivalent up to constant
depth and polynomial size \cite{GHMP:fanout,Moore:fanout}.  That is,
each gate above can be simulated exactly by a constant-depth,
polynomial-size quantum circuit using any of the other gates above,
together with standard one- and two-qubit gates (e.g., $\CNOT$, $H$,
and $T$).  This is not true in the classical case.  Furthermore, using
fanout gates, in constant depth and polynomial size one can
approximate $n$-qubit threshold gates, unbounded AND (generalized
Toffoli) gates and OR gates, sorting, arithmetical operations, phase
estimation, and the quantum Fourier transform
\cite{HS:fanout,Spalek:fanout}.  Since long quantum computations may
be difficult to maintain due to decoherence, shallow quantum circuits
may prove much more realistic, at least in the short term.

On the negative side, fanout gates so far appear hard to implement.
There is recent theoretical evidence that fanout gates cannot be
simulated in small depth and small width, even if unbounded AND gates
are allowed \cite{FGHZ:ancillae}.\footnote{Any circuit of depth $d$
using AND and single-qubit gates to compute $n$-qubit fanout provably
needs at least $n/2^d$ ancill\ae.}  All these results underscore how
crucial fanout gates are to providing powerful, small-depth quantum
computation.

Rather than trying to implement fanout with a traditional small-depth
quantum circuit, we can perhaps take another promising approach:
evolve an $n$-qubit system according to a (hopefully implementable)
Hamiltonian.  We show that a simple Hamiltonian, similar to one
suggested recently by Chuang \cite{Chuang:hamiltonian}, does exactly
this.

Let $\Ztot{n}$ be the $n$-qubit operator $\frac{1}{2}\sum_{i=1}^n
Z_i$, where $Z_i$ is the $Z$-gate acting on the $i$th qubit.  If, for
instance, each qubit is represented by a spin-$\frac{1}{2}$ particle,
then $\Ztot{n}$ is the observable representing the total spin angular
momentum in the $z$-direction.  Let $J>0$ be some constant in units of
energy.  We will show that for even $n$, the $(n+1)$-qubit fanout gate
arises naturally by evolving the first $n$ qubits through the
Hamiltonian
\[ H_n \eqdf J(\Ztot{n})^2\]
twice---first for time $\frac{\pi\hbar}{2J}$ then for time
$\frac{3\pi\hbar}{2J}$---together with a modest amount of additional
processing.  We also show that evolving the $3$-qubit Hamiltonian
$J(\Ztot{3})^2$ results (modulo a global phase factor) in a $3$-qubit
``inversion on equality'' gate $I_=$, which maps $\ket{abc}$ to
$(-1)^{\delta_{ab}\delta_{ac}}\ket{abc}$.  Applying $I_=$ with one of
the three qubits set to $\ket{1}$ results in a controlled $Z$-gate on
the other two qubits, which is easily converted to a $\CNOT$ gate.

Chuang's proposed Hamiltonian is closely related to $H_n$.  It is
\[ K_n \eqdf \sum_{1\leq i < j \leq n} J_{i,j} Z_i Z_j, \]
where the $J_{i,j}$ are energy coefficients, and may be potentially
realizable for certain combinations of the $J_{i,j}$, depending on the
physical arrangement of the qubits \cite{Chuang:hamiltonian}.  $K_n$
is the sum of pairwise interactions between the particles.  In the
special case where all the $J_{i,j}$ are equal to $J/2$, we see that
$K_n$ differs from $H_n$ by a multiple of the identity, and so
evolving through $H_n$ and evolving through $K_n$ are equivalent up to
an overall phase factor.

We will show the $I_=$-gate in Section~\ref{sec:eq-gate}.  In
Section~\ref{sec:parity-gate} we show the implementation of $F_{n+1}$
for even $n$.  In the sequel, we will assume for convenience that
$\hbar = J/2 = 1$.  If $X$ and $Y$ are vectors or operators, we say,
``$X\propto Y$'' to mean that $X = e^{i\theta}Y$ for some real
$\theta$, that is, $X = Y$ up to an overall phase factor.  We use the
same notation with individual components of $X$ and $Y$, meaning that
the phase factor is independent of which component we choose.  If $A$
is a set, we let $\card{A}$ denote the cardinality of $A$.

\section{Main Results}

\subsection{\boldmath Evolving $H_n$}

Since $H_n$ is represented in the computational basis by a diagonal
matrix, it is particularly easy to see how it evolves in time.  We
only need to find the value of each diagonal element.  Let $\vec{x} =
x_1\cdots x_n$ be a vector of $n$ bits.  For $1 \leq i,j \leq n$, we
see that $Z_iZ_j\ket{\vec{x}} = (-1)^{x_i\oplus x_j}\ket{\vec{x}}$,
and thus $Z_iZ_j$ flips the sign iff $x_i \neq x_j$.  Suppose $k$ of
the $x_i$ are $1$ and the rest ($n-k$) are $0$.  Then,
\begin{eqnarray*}
\braopket{\vec{x}}{H_n}{\vec{x}} & = & \frac{1}{2} \sum_{i=1}^n
\sum_{j=1}^n (-1)^{x_i\oplus x_j} \\
& = & (1/2) \left( \card{\setof{(i,j) \condition x_i = x_j}} -
\card{\setof{(i,j) \condition x_i \neq x_j}} \right) \\
& = & (1/2)(k^2 + (n-k)^2 - 2k(n-k)) \\
& = & (1/2)(n^2 + 4k^2 - 4nk) \\
& = & n^2/2 - 2k(n-k).
\end{eqnarray*}
The $n^2/2$ term is independent of $\vec{x}$, so up to addition of a
multiple of $I$, the diagonal term is effectively $-2k(n-k)$.
We evolve $H_n$ for time $t = \pi/4$.  Let $U_n \eqdf e^{-iH_nt}$
where $t = \pi/4$.  We get
\begin{equation}\label{eqn:diagonal-elt}
\braopket{\vec{x}}{U_n}{\vec{x}} \propto e^{i\pi k(n-k)/2} =
i^{k(n-k)}.
\end{equation}
All the off-diagonal matrix elements of $U_n$ are zero.

\subsection{\boldmath Implementing the $I_=$-Gate}
\label{sec:eq-gate}

For the $n = 3$ case, (\ref{eqn:diagonal-elt}) yields
\[ U_3 \propto \diag{rrrrrrrr}{1 & -1 & -1 & -1 & -1 & -1 & -1 & 1}
\propto I_=. \]
No further gates are required.

\subsection{\boldmath Implementing the $F_{n+1}$-Gate}
\label{sec:parity-gate}

\begin{figure}
\begin{center}
\begin{picture}(0,0)%
\includegraphics{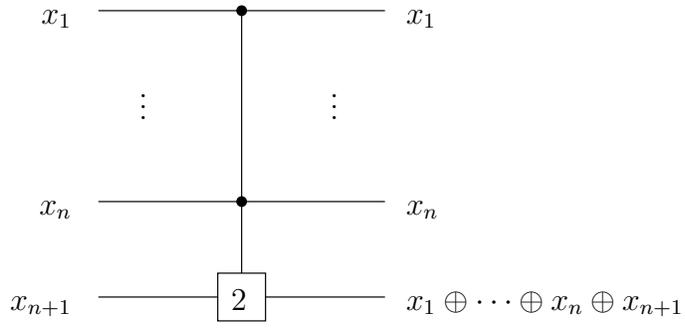}%
\end{picture}%
\setlength{\unitlength}{3947sp}%
\begingroup\makeatletter\ifx\SetFigFont\undefined%
\gdef\SetFigFont#1#2#3#4#5{%
  \reset@font\fontsize{#1}{#2pt}%
  \fontfamily{#3}\fontseries{#4}\fontshape{#5}%
  \selectfont}%
\fi\endgroup%
\begin{picture}(2100,2082)(2251,-2923)
\put(2701,-1636){\makebox(0,0)[b]{\smash{\SetFigFont{12}{14.4}{\rmdefault}{\mddefault}{\updefault}$\vdots$}}}
\put(3901,-1636){\makebox(0,0)[b]{\smash{\SetFigFont{12}{14.4}{\rmdefault}{\mddefault}{\updefault}$\vdots$}}}
\put(2251,-2236){\makebox(0,0)[rb]{\smash{\SetFigFont{12}{14.4}{\rmdefault}{\mddefault}{\updefault}$x_n$}}}
\put(4351,-2236){\makebox(0,0)[lb]{\smash{\SetFigFont{12}{14.4}{\rmdefault}{\mddefault}{\updefault}$x_n$}}}
\put(4351,-2836){\makebox(0,0)[lb]{\smash{\SetFigFont{12}{14.4}{\rmdefault}{\mddefault}{\updefault}$x_1\oplus \cdots \oplus x_n \oplus x_{n+1}$}}}
\put(4351,-1036){\makebox(0,0)[lb]{\smash{\SetFigFont{12}{14.4}{\rmdefault}{\mddefault}{\updefault}$x_1$}}}
\put(2251,-1036){\makebox(0,0)[rb]{\smash{\SetFigFont{12}{14.4}{\rmdefault}{\mddefault}{\updefault}$x_1$}}}
\put(2251,-2836){\makebox(0,0)[rb]{\smash{\SetFigFont{12}{14.4}{\rmdefault}{\mddefault}{\updefault}$x_{n+1}$}}}
\put(3301,-2836){\makebox(0,0)[b]{\smash{\SetFigFont{12}{14.4}{\rmdefault}{\mddefault}{\updefault}$2$}}}
\end{picture}
\caption{Definition of the parity gate.}\label{fig:parity-gate}
\end{center}
\end{figure}

\begin{figure}
\begin{center}
\begin{picture}(0,0)%
\includegraphics{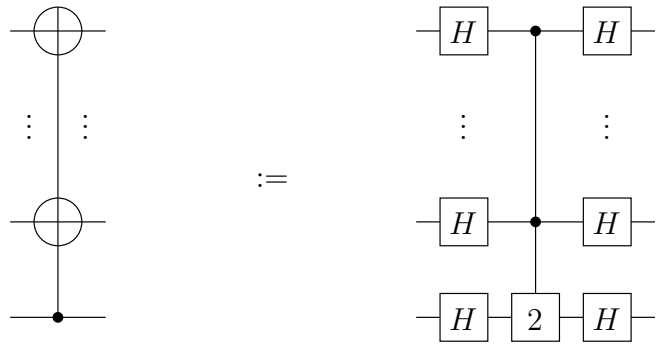}%
\end{picture}%
\setlength{\unitlength}{3947sp}%
\begingroup\makeatletter\ifx\SetFigFont\undefined%
\gdef\SetFigFont#1#2#3#4#5{%
  \reset@font\fontsize{#1}{#2pt}%
  \fontfamily{#3}\fontseries{#4}\fontshape{#5}%
  \selectfont}%
\fi\endgroup%
\begin{picture}(4074,2124)(3889,-2623)
\put(5551,-1636){\makebox(0,0)[b]{\smash{\SetFigFont{12}{14.4}{\rmdefault}{\mddefault}{\updefault}$:=$}}}
\put(4051,-1336){\makebox(0,0)[rb]{\smash{\SetFigFont{12}{14.4}{\rmdefault}{\mddefault}{\updefault}$\vdots$}}}
\put(4351,-1336){\makebox(0,0)[lb]{\smash{\SetFigFont{12}{14.4}{\rmdefault}{\mddefault}{\updefault}$\vdots$}}}
\put(6751,-1336){\makebox(0,0)[b]{\smash{\SetFigFont{12}{14.4}{\rmdefault}{\mddefault}{\updefault}$\vdots$}}}
\put(7651,-1336){\makebox(0,0)[b]{\smash{\SetFigFont{12}{14.4}{\rmdefault}{\mddefault}{\updefault}$\vdots$}}}
\put(6751,-736){\makebox(0,0)[b]{\smash{\SetFigFont{12}{14.4}{\rmdefault}{\mddefault}{\updefault}$H$}}}
\put(6751,-1936){\makebox(0,0)[b]{\smash{\SetFigFont{12}{14.4}{\rmdefault}{\mddefault}{\updefault}$H$}}}
\put(6751,-2536){\makebox(0,0)[b]{\smash{\SetFigFont{12}{14.4}{\rmdefault}{\mddefault}{\updefault}$H$}}}
\put(7651,-736){\makebox(0,0)[b]{\smash{\SetFigFont{12}{14.4}{\rmdefault}{\mddefault}{\updefault}$H$}}}
\put(7651,-1936){\makebox(0,0)[b]{\smash{\SetFigFont{12}{14.4}{\rmdefault}{\mddefault}{\updefault}$H$}}}
\put(7651,-2536){\makebox(0,0)[b]{\smash{\SetFigFont{12}{14.4}{\rmdefault}{\mddefault}{\updefault}$H$}}}
\put(7201,-2536){\makebox(0,0)[b]{\smash{\SetFigFont{12}{14.4}{\rmdefault}{\mddefault}{\updefault}$2$}}}
\end{picture}
\caption{Implementing fanout by a parity gate and Hadamard
gates.}\label{fig:parity-to-fanout}
\end{center}
\end{figure}

Let $n$ be even.  $U_n$ is actually more closely related to the
$(n+1)$-qubit (classical) \emph{parity gate}, shown in
Figure~\ref{fig:parity-gate}.
We will show how to implement parity with $U_n$.  This suffices,
because the equality given in Figure~\ref{fig:parity-to-fanout}
was observed in \cite{Moore:fanout,GHMP:fanout}, where $H$ is the
Hadamard gate.

\begin{figure}
\begin{center}
\begin{picture}(0,0)%
\includegraphics{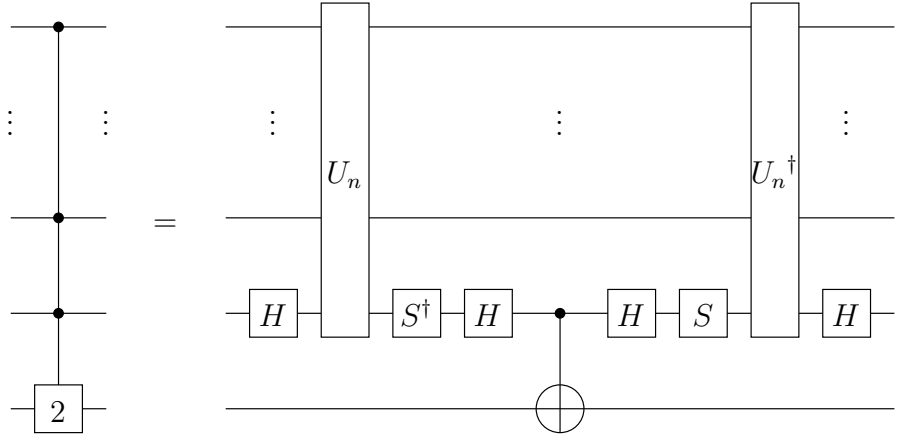}%
\end{picture}%
\setlength{\unitlength}{3947sp}%
\begingroup\makeatletter\ifx\SetFigFont\undefined%
\gdef\SetFigFont#1#2#3#4#5{%
  \reset@font\fontsize{#1}{#2pt}%
  \fontfamily{#3}\fontseries{#4}\fontshape{#5}%
  \selectfont}%
\fi\endgroup%
\begin{picture}(5574,2724)(4489,-3223)
\put(6151,-1336){\makebox(0,0)[b]{\smash{\SetFigFont{12}{14.4}{\rmdefault}{\mddefault}{\updefault}$\vdots$}}}
\put(9751,-1336){\makebox(0,0)[b]{\smash{\SetFigFont{12}{14.4}{\rmdefault}{\mddefault}{\updefault}$\vdots$}}}
\put(7951,-1336){\makebox(0,0)[b]{\smash{\SetFigFont{12}{14.4}{\rmdefault}{\mddefault}{\updefault}$\vdots$}}}
\put(5101,-1336){\makebox(0,0)[b]{\smash{\SetFigFont{12}{14.4}{\rmdefault}{\mddefault}{\updefault}$\vdots$}}}
\put(4501,-1336){\makebox(0,0)[b]{\smash{\SetFigFont{12}{14.4}{\rmdefault}{\mddefault}{\updefault}$\vdots$}}}
\put(5476,-1936){\makebox(0,0)[b]{\smash{\SetFigFont{12}{14.4}{\rmdefault}{\mddefault}{\updefault}$=$}}}
\put(6151,-2536){\makebox(0,0)[b]{\smash{\SetFigFont{12}{14.4}{\rmdefault}{\mddefault}{\updefault}$H$}}}
\put(7051,-2536){\makebox(0,0)[b]{\smash{\SetFigFont{12}{14.4}{\rmdefault}{\mddefault}{\updefault}$\adj{S}$}}}
\put(7501,-2536){\makebox(0,0)[b]{\smash{\SetFigFont{12}{14.4}{\rmdefault}{\mddefault}{\updefault}$H$}}}
\put(8401,-2536){\makebox(0,0)[b]{\smash{\SetFigFont{12}{14.4}{\rmdefault}{\mddefault}{\updefault}$H$}}}
\put(8851,-2536){\makebox(0,0)[b]{\smash{\SetFigFont{12}{14.4}{\rmdefault}{\mddefault}{\updefault}$S$}}}
\put(9751,-2536){\makebox(0,0)[b]{\smash{\SetFigFont{12}{14.4}{\rmdefault}{\mddefault}{\updefault}$H$}}}
\put(6601,-1636){\makebox(0,0)[b]{\smash{\SetFigFont{12}{14.4}{\rmdefault}{\mddefault}{\updefault}$U_n$}}}
\put(9301,-1636){\makebox(0,0)[b]{\smash{\SetFigFont{12}{14.4}{\rmdefault}{\mddefault}{\updefault}$\adj{U_n}$}}}
\put(4801,-3136){\makebox(0,0)[b]{\smash{\SetFigFont{12}{14.4}{\rmdefault}{\mddefault}{\updefault}$2$}}}
\end{picture}
\caption{$U_n$ implementing the parity gate.}\label{fig:final-circuit}
\end{center}
\end{figure}

\begin{figure}
\begin{center}
\begin{picture}(0,0)%
\includegraphics{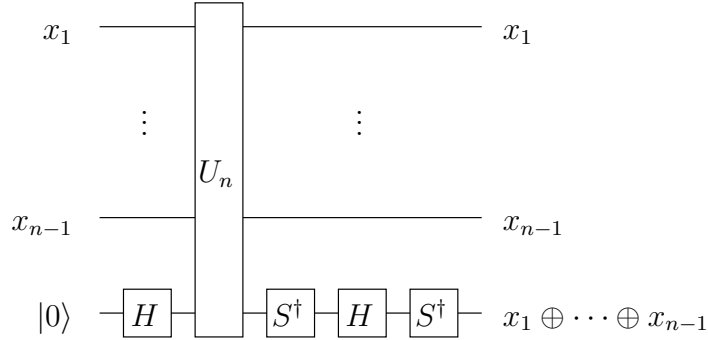}%
\end{picture}%
\setlength{\unitlength}{3947sp}%
\begingroup\makeatletter\ifx\SetFigFont\undefined%
\gdef\SetFigFont#1#2#3#4#5{%
  \reset@font\fontsize{#1}{#2pt}%
  \fontfamily{#3}\fontseries{#4}\fontshape{#5}%
  \selectfont}%
\fi\endgroup%
\begin{picture}(2700,2124)(5701,-2623)
\put(6151,-1336){\makebox(0,0)[b]{\smash{\SetFigFont{12}{14.4}{\rmdefault}{\mddefault}{\updefault}$\vdots$}}}
\put(7501,-1336){\makebox(0,0)[b]{\smash{\SetFigFont{12}{14.4}{\rmdefault}{\mddefault}{\updefault}$\vdots$}}}
\put(5701,-736){\makebox(0,0)[rb]{\smash{\SetFigFont{12}{14.4}{\rmdefault}{\mddefault}{\updefault}$x_1$}}}
\put(5701,-1936){\makebox(0,0)[rb]{\smash{\SetFigFont{12}{14.4}{\rmdefault}{\mddefault}{\updefault}$x_{n-1}$}}}
\put(5701,-2536){\makebox(0,0)[rb]{\smash{\SetFigFont{12}{14.4}{\rmdefault}{\mddefault}{\updefault}$\ket{0}$}}}
\put(8401,-736){\makebox(0,0)[lb]{\smash{\SetFigFont{12}{14.4}{\rmdefault}{\mddefault}{\updefault}$x_1$}}}
\put(8401,-1936){\makebox(0,0)[lb]{\smash{\SetFigFont{12}{14.4}{\rmdefault}{\mddefault}{\updefault}$x_{n-1}$}}}
\put(8401,-2536){\makebox(0,0)[lb]{\smash{\SetFigFont{12}{14.4}{\rmdefault}{\mddefault}{\updefault}$x_1 \oplus \cdots \oplus x_{n-1}$}}}
\put(6151,-2536){\makebox(0,0)[b]{\smash{\SetFigFont{12}{14.4}{\rmdefault}{\mddefault}{\updefault}$H$}}}
\put(7051,-2536){\makebox(0,0)[b]{\smash{\SetFigFont{12}{14.4}{\rmdefault}{\mddefault}{\updefault}$\adj{S}$}}}
\put(7501,-2536){\makebox(0,0)[b]{\smash{\SetFigFont{12}{14.4}{\rmdefault}{\mddefault}{\updefault}$H$}}}
\put(7951,-2536){\makebox(0,0)[b]{\smash{\SetFigFont{12}{14.4}{\rmdefault}{\mddefault}{\updefault}$\adj{S}$}}}
\put(6601,-1636){\makebox(0,0)[b]{\smash{\SetFigFont{12}{14.4}{\rmdefault}{\mddefault}{\updefault}$U_n$}}}
\end{picture}
\caption{A circuit similar to the parity gate.}\label{fig:parity-like}
\end{center}
\end{figure}

Suppose for now that $n \equiv 2 \pmod 4$, and that $\vec{x}$ has $k$
ones and $n-k$ zeros.  If $k$ is even, then so is $n-k$, and thus
$k(n-k) \equiv 0 \pmod 4$.  If $k$ is odd, then clearly, $k(n-k)
\equiv 1 \pmod 4$.  Thus we have, by (\ref{eqn:diagonal-elt}),
\begin{equation}\label{eqn:n-2-mod-4}
\braopket{\vec{x}}{U_n}{\vec{x}} \propto \left\{ \begin{array}{ll}
1 & \mbox{if $\vec{x}$ has even parity,} \\
i & \mbox{if $\vec{x}$ has odd parity.}
\end{array} \right.
\end{equation}
We start with the $(n+1)$-qubit basis state $\ket{x_1\cdots
x_{n-1}rb}$, where $x_1,\ldots,x_{n-1},r,b\in\two$.  We prepare the
$n$th qubit in the state
\[ H\ket{r} = (\ket{0} + (-1)^r\ket{1})/\sqrt{2}, \]
so that the current state of the first $n$ qubits is $\ket{x_1\cdots
x_{n-1}}(\ket{0} + (-1)^r\ket{1})/\sqrt{2}$.  If we run this state through
$U_n$, then by (\ref{eqn:n-2-mod-4}) the result is
\[ \frac{1}{\sqrt{2}}\ket{x_1\cdots x_{n-1}}(i^p\ket{0} +
i^{1-p}(-1)^r\ket{1}), \]
where $p = (x_1 + \cdots + x_{n-1})\bmod 2$.  The two states of the
$n$th qubit corresponding to the two values of $p$ are
orthogonal, so we just need to rotate the $n$th qubit back to the
computational basis.  The $n$th qubit state is either $+y$ or $-y$ on
the Bloch sphere, depending on $p$ and $r$, so we can use
$H\adj{S}$, where S is the phase gate $\diag{rr}{1 & i}$, to rotate
the $y$-axis to the $z$-axis.  The final circuit is shown in
Figure~\ref{fig:final-circuit}.
When the $\CNOT$ gate is applied, its control qubit can be seen to be
in the state $i^p\ket{p \oplus r}$, unentangled with the other qubits.
The rest of the circuit is then needed to uncompute the conditional
phase factor $i^p$.  We can implement $\adj{U_n}$ by evolving $-H_n$
for time $\pi/4$, or equivalently, by evolving $H_n$ for time
$3\pi/4$, since $U_n^4 = I$.  Of course, if we are willing to keep the
phase factor, we can get by without this part of the circuit.  In
fact, we can get the parity-like gate of Figure~\ref{fig:parity-like}
with only one use of $U_n$.

\begin{figure}
\begin{center}
\begin{picture}(0,0)%
\includegraphics{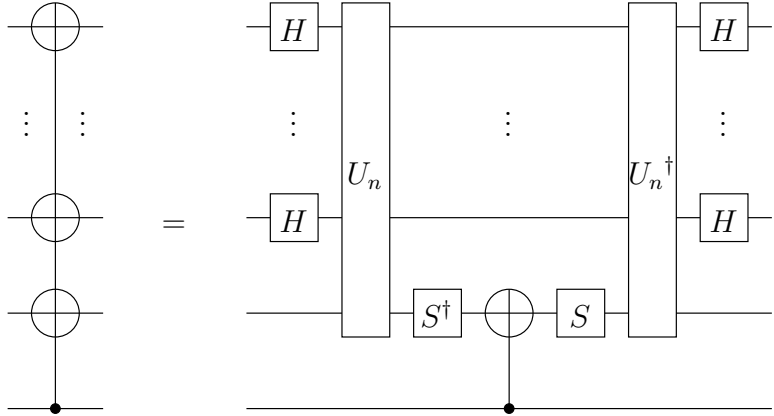}%
\end{picture}%
\setlength{\unitlength}{3947sp}%
\begingroup\makeatletter\ifx\SetFigFont\undefined%
\gdef\SetFigFont#1#2#3#4#5{%
  \reset@font\fontsize{#1}{#2pt}%
  \fontfamily{#3}\fontseries{#4}\fontshape{#5}%
  \selectfont}%
\fi\endgroup%
\begin{picture}(4824,2599)(4339,-3098)
\put(4501,-1336){\makebox(0,0)[rb]{\smash{\SetFigFont{12}{14.4}{\rmdefault}{\mddefault}{\updefault}$\vdots$}}}
\put(4801,-1336){\makebox(0,0)[lb]{\smash{\SetFigFont{12}{14.4}{\rmdefault}{\mddefault}{\updefault}$\vdots$}}}
\put(6151,-1336){\makebox(0,0)[b]{\smash{\SetFigFont{12}{14.4}{\rmdefault}{\mddefault}{\updefault}$\vdots$}}}
\put(7501,-1336){\makebox(0,0)[b]{\smash{\SetFigFont{12}{14.4}{\rmdefault}{\mddefault}{\updefault}$\vdots$}}}
\put(8851,-1336){\makebox(0,0)[b]{\smash{\SetFigFont{12}{14.4}{\rmdefault}{\mddefault}{\updefault}$\vdots$}}}
\put(5401,-1936){\makebox(0,0)[b]{\smash{\SetFigFont{12}{14.4}{\rmdefault}{\mddefault}{\updefault}$=$}}}
\put(7051,-2536){\makebox(0,0)[b]{\smash{\SetFigFont{12}{14.4}{\rmdefault}{\mddefault}{\updefault}$\adj{S}$}}}
\put(6151,-736){\makebox(0,0)[b]{\smash{\SetFigFont{12}{14.4}{\rmdefault}{\mddefault}{\updefault}$H$}}}
\put(6151,-1936){\makebox(0,0)[b]{\smash{\SetFigFont{12}{14.4}{\rmdefault}{\mddefault}{\updefault}$H$}}}
\put(7951,-2536){\makebox(0,0)[b]{\smash{\SetFigFont{12}{14.4}{\rmdefault}{\mddefault}{\updefault}$S$}}}
\put(8851,-736){\makebox(0,0)[b]{\smash{\SetFigFont{12}{14.4}{\rmdefault}{\mddefault}{\updefault}$H$}}}
\put(8851,-1936){\makebox(0,0)[b]{\smash{\SetFigFont{12}{14.4}{\rmdefault}{\mddefault}{\updefault}$H$}}}
\put(6601,-1636){\makebox(0,0)[b]{\smash{\SetFigFont{12}{14.4}{\rmdefault}{\mddefault}{\updefault}$U_n$}}}
\put(8401,-1636){\makebox(0,0)[b]{\smash{\SetFigFont{12}{14.4}{\rmdefault}{\mddefault}{\updefault}$\adj{U_n}$}}}
\end{picture}
\caption{$U_n$ implementing the fanout
gate.}\label{fig:fanout-implementation}
\end{center}
\end{figure}

As we mentioned before, one gets a fanout gate by applying Hadamards
on each qubit on both sides of a parity gate.  So starting with the
circuit in Figure~\ref{fig:final-circuit}, after some
simplification we get the circuit shown in
Figure~\ref{fig:fanout-implementation}.
If $n \equiv 0 \pmod 4$, then we get a similar analysis of $U_n$,
except that (\ref{eqn:n-2-mod-4}) becomes
\[ \braopket{\vec{x}}{U_n}{\vec{x}} \propto \left\{ \begin{array}{ll}
1 & \mbox{if $\vec{x}$ has even parity,} \\
-i & \mbox{if $\vec{x}$ has odd parity.}
\end{array} \right. \]
It follows that we can swap $U_n$ with $\adj{U_n}$ in
Figures~\ref{fig:final-circuit}, \ref{fig:parity-like}, and
\ref{fig:fanout-implementation} above to maintain the equalities.

\section{Conclusions and Further Research}

A key point in our implementation is that the number of terms in the
Hamiltonian $H_n$ is quadratic in $n$, which gives a quadratic term in
the phase shift.  We suspect there is also some way to get parity from
the more general $K_n$ when the $J_{i,j}$ are not all equal, provided
there are still quadratically many terms.  We also suspect that this
will not work where there are fewer than quadratically many terms, for
example, in the case where we just consider interactions between
adjacent paricles in a ring, i.e., $J_{i,j} = J > 0$ if $j \equiv i+1
\pmod n$, and $J_{i,j} = 0$ otherwise.  In this case, there are only
linearly many terms in the Hamiltonian.

A second Hamiltonian described by Chuang as potentially realizable in
the lab \cite{Chuang:hamiltonian} is
\[ L_n = \sum_{1\leq i < j \leq n} J_{i,j}(X_iX_j + Y_iY_j + Z_iZ_j),
\]
which, when all the $J_{i,j} = 2$, differs from the squared total spin
$L^2 \eqdf (\Xtot{n})^2 + (\Ytot{n})^2 + (\Ztot{n})^2$ by a multiple
of $I$.  We conjecture that $L^2$ can also be used to implement
parity.


\end{document}